\documentclass[%
 aip,
 apl,
 amsmath,amssymb,
 reprint,%
]{revtex4-1}

\usepackage{graphicx}
\usepackage{graphicx}
\graphicspath{ {./figures/} }
\usepackage{dcolumn}
\usepackage{bm}

\usepackage[utf8]{inputenc}
\usepackage[T1]{fontenc}
\usepackage{mathptmx}

\usepackage[binary-units=true]{siunitx}
\usepackage{textgreek}

\newcommand\db{$\delta/\beta$}
\newcommand\sig{$\sigma$}

\begin{document}


\title{Optimizing contrast and spatial resolution in hard X-ray tomography of medically relevant tissues}


\author{Griffin Rodgers}
\author{Georg Schulz}
\affiliation{Biomaterials Science Center, University of Basel, 4123 Allschwil, Switzerland}

\author{Hans Deyhle}
\affiliation{Biomaterials Science Center, University of Basel, 4123 Allschwil, Switzerland}
\affiliation{Diamond Light Source Ltd., Didcot OX11 0DE, UK}%

\author{Willy Kuo}
\affiliation{Biomaterials Science Center, University of Basel, 4123 Allschwil, Switzerland}
\affiliation{The Interface Group, Institute of Physiology, University of Zurich, 8057 Zurich, Switzerland}

\author{Christoph Rau}
\affiliation{Diamond Light Source Ltd., Didcot OX11 0DE, UK}%

\author{Timm Weitkamp}
\email{weitkamp@synchrotron-soleil.fr}
\affiliation{Synchrotron Soleil, 91192 Gif-sur-Yvette, France}

\author{Bert M\"uller}
\affiliation{Biomaterials Science Center, University of Basel, 4123 Allschwil, Switzerland}

\date{\today}

\begin{abstract}
Hard X-ray tomography with Paganin’s widespread single-distance phase retrieval filter improves contrast-to-noise ratio (CNR) while reducing spatial resolution (SR). We demonstrate that a Gaussian filter provided larger CNR at high SR with interpretable density measurements for two medically relevant soft tissue samples. Paganin's filter produced larger CNR at low SR, though \emph{a priori} assumptions were generally false and image quality gains diminish for CNR $>1$. Therefore, simple absorption measurements of low-$Z$ specimens combined with Gaussian filtering can provide improved image quality and model-independent density measurements compared to single-distance phase retrieval.
\end{abstract}


\maketitle


Hard X-ray micro tomography non-destructively provides a three-dimensional map of a wide range of samples.\cite{stock2008microcomputed} Medically relevant tissues are composed primarily of low-$Z$ elements and therefore have low X-ray absorption. Phase contrast methods have been proposed because the phase cross-section is orders of magnitude larger than the absorption cross-section for photon energies on the order of \SI{10}{\kilo\electronvolt}.\cite{momose2005recent}

The simplest phase contrast techniques are propagation-based, where Fresnel diffraction from free-space propagation of the X-ray wavefront makes phase information detectable in intensity measurements.\cite{paganin2006coherent} Sufficient coherence and propagation distance allow for the wavefront phase shift to be extracted from radiographs at one or more positions downstream.\cite{cloetens1999holotomography,bronnikov1999reconstruction} Exact phase retrieval for samples with non-negligible absorption may require several downstream positions or multiple energies.\cite{gureyev2001quantitative} Still, single-distance approaches can provide semi-quantitative phase retrieval, reduce edge enhancement, and in some cases improve the contrast-to-noise ratio (CNR).\cite{boone2012improved} The single-distance phase retrieval filter introduced by Paganin \emph{et al.}\ is among the most used thanks to its simplicity and large CNR gains.\cite{paganin2002simultaneous} This method has found many applications, for example low-dose medical imaging and virtual histology.\cite{arfelli1998low,topperwien2018three,albers2018histology,hieber2016nucleolar}

However, image quality is a function of both the CNR and the spatial resolution (SR),\cite{grodzins1983optimum,thurner2004optimization} which Paganin filtering reduces.\cite{boone2013thesis} It is often unclear if CNR gains should be attributed to phase sensitivity or the low-pass effects. Previous experiments have shown that pixel binning of tomography measurements improve the CNR at the expense of SR.\cite{thurner2004optimization,thalmann2017single} The question now arises if low-pass filtering of absorption datasets\cite{bikis2019sensitivity} outperforms Paganin phase retrieval when both CNR and SR are taken into account.

In this study, we present two synchrotron radiation-based microtomography (SR\textmu CT) measurements of biological tissues: a cylinder of paraffin-embedded human cerebellum and a formalin-fixed mouse kidney. Gaussian filtering was applied to the projections prior to reconstruction in absorption contrast mode, while Paganin filtering was used for phase retrieval of the projections before reconstruction. The CNR and SR were measured for a range of filter kernels to optimize image quality.

The human cerebellum tissue was extracted post-mortem with informed consent for scientific use and embedded according to the standard procedure for pathological analysis, i.e.\ transferred to 4\% histological-grade buffered paraformaldehyde, ascending ethanol solutions, xylene, and finally embedded in a paraffin mixture. A stainless-steel punch was used to produce the final \SI{6}{\milli\meter} diameter sample.

One seven-month old female C57BL/6J mouse was anaesthetized with Ketamine/Xylazine and the kidneys were perfused retrogradely via the abdominal aorta\cite{czogalla2016mouse} at 150 mmHg with 10 ml of phosphate-buffered saline (PBS), 100 ml 4\% formaldehyde/1\% glutaraldehyde/PBS, 20 ml PBS, 50 ml glycine solution (5 mg/ml in PBS), 40 ml PBS. 4 ml of X-ray contrast agent solution (75 mg iodine/ml) was injected into the vasculature to distinguish blood vessels from tubules. The abdominal cavity was then filled with 4\% glutaraldehyde/PBS to crosslink the contrast agent, and the kidneys excised and kept in 4\% glutaraldehyde/PBS. The right kidney was mounted in 1\% Agar/PBS in a 0.5 ml plastic tubes intended for polymerase chain reaction (PCR tubes) to avoid movement during scanning.

The SR\textmu CT measurements were performed at Diamond Manchester Imaging Beamline (I13-2, Diamond Light Source, Didcot, UK), where an undulator source is used.\cite{rau2011coherent} Table \ref{tab_acq_params} shows the acquisition parameters for both measurements.

\begin{table}[t]
\caption{\label{tab_acq_params} Summary of the acquisition parameters for the human cerebellum and mouse kidney samples measured at the Diamond Manchester Imaging Beamline (I13-2, Diamond Light Source, Didcot, UK). $z_c$ is the critical distance for the single distance phase retrieval as defined by Weitkamp \emph{et al.}\cite{weitkamp2011ankaphase}}
\begin{ruledtabular}
\begin{tabular}{l c c}
Sample & Brain & Kidney \\ 
\hline  \noalign{\smallskip}
Mode & monochromatic & filtered pink beam\\
Mean photon energy & \SI{20}{\kilo\electronvolt} & \SI{23}{\kilo\electronvolt} \\
Camera & pco.4000\footnote{PCO AG, Kelheim, Germany} & pco.4000 \\
Bit depth & 14 & 14 \\
Objective &  PLAPON 2X\footnote{Olympus Corporation, Tokyo, Japan} &  PLAPON 4X \\
Numerical aperture & $0.08$ & $0.16$ \\
Hardware binning & $2\times2$ & $1\times1$ \\
Effective pixel size & \SI{4.6}{\micro\meter} & \SI{1.125}{\micro\meter} \\
Scintillator & LuAG & LuAG \\
		    & \SI{500}{\micro\meter} & \SI{500}{\micro\meter} \\
Detector distance & \SI{7}{\centi\meter} & \SI{9}{\centi\meter} \\
$z_c$ & \SI{137}{\centi\meter} & \SI{9.4}{\centi\meter} \\
Sample transmission\footnote{Mean over ROI in sample center.}
 & 75\% & 56\% \\
Acquisition mode & standard & off-axis \\
				& step scan & fly scan \\
Number of projections & 1201 & 2501 \\
Exposure time & 2 s & 0.5 s \\
Mean flat-field counts & 10,000 & 1,700 \\
\end{tabular}
\end{ruledtabular}
\end{table}

\begin{figure}[t] 
\includegraphics[scale=0.95]{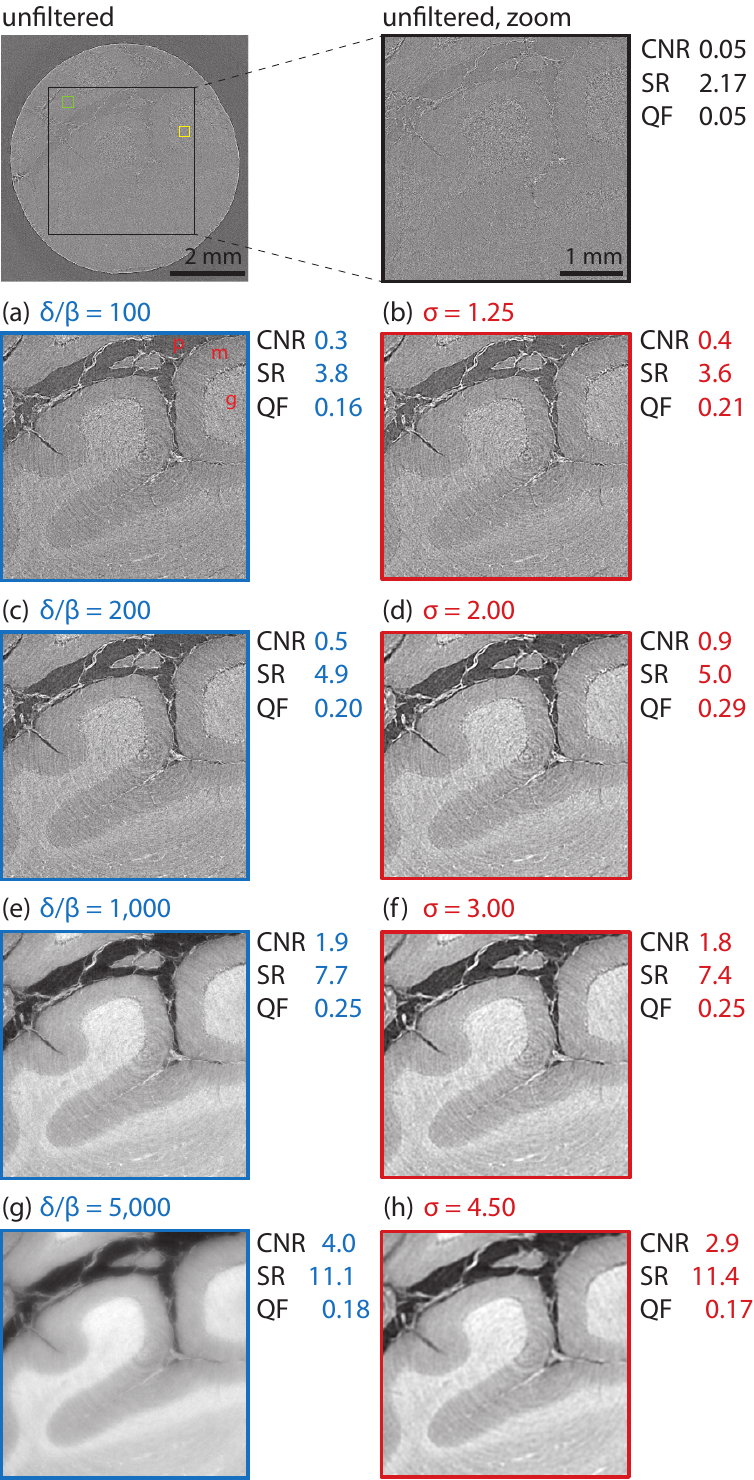}%
\caption{\label{fig_slices_brain} Reconstructed slices of the human cerebellum after Paganin phase retrieval (left, blue) and Gaussian filtering (right, red) of the transmission projections. Filter kernel, SR (in pixels), CNR, and QF are indicated. Regions for CNR measurement are given in the unfiltered slice (green and yellow). Grayscale is given by the intersection of the histogram of each zoom with a threshold.}
\end{figure}

The optimal photon energy criteria for an absorption contrast measurement is given by $\mu(E)\times D=2$ (equivalent to 13\% transmission), where $\mu(E)$ is the linear attenuation coefficient and $D$ the sample diameter.\cite{grodzins1983optimum} Both the brain and kidney samples have $\mu \times D<2$, indicating higher than optimal photon energies. 

\begin{figure*}[t]  
\includegraphics[scale=0.95]{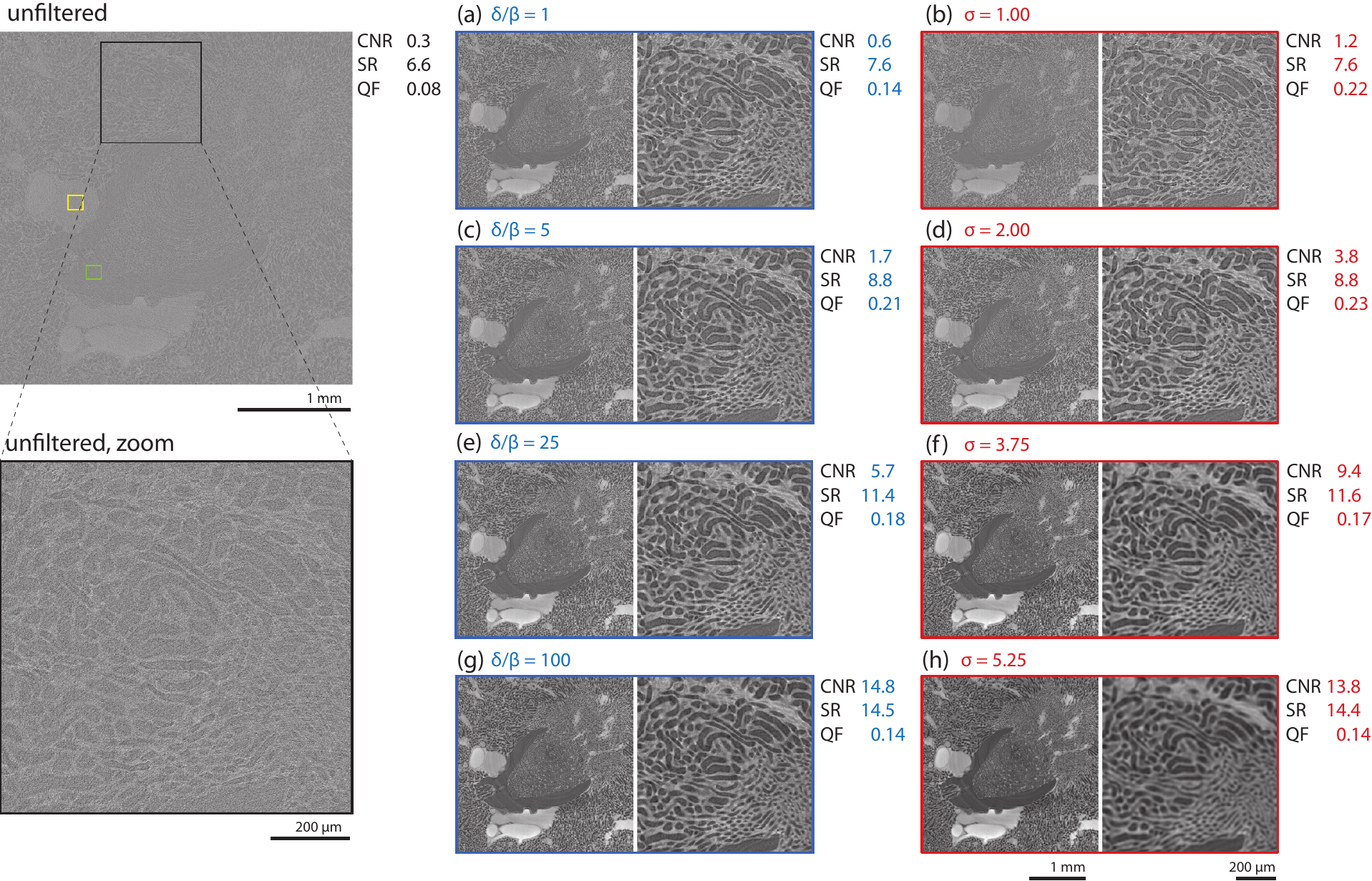}%
\caption{\label{fig_slices_kidney} Reconstructed slices of the mouse kidney after Paganin phase retrieval (left, blue) and Gaussian filtering (right, red) of transmission projections. Filter kernel, CNR, SR (pixels), and QF are indicated. Zooms help to identify reduced SR and improved CNR with increasing filter size. Regions for CNR measurement are given in the unfiltered slice (green and yellow). Grayscale based on the intersection of the histogram of each zoom with a threshold.}%
\end{figure*}

The conditions for validity of the Paganin phase retrieval are (i) single homogenous material or constant \db\ throughout the specimen, (ii) monochromatic plane wave, and (iii) the detector is in the near-field.\cite{weitkamp2011ankaphase} It should be noted negligible sample absorption is not required. Previous X-ray grating interferometry measurements of the brain sample confirm constant \db\ ratio,\cite{bikis2019sensitivity} however the \db\ ratio is unknown for the mouse kidney. Condition (ii) is valid for the brain measurement with a double-crystal monochromator and previous experimental results have shown Paganin's method is robust for polychromaticity such as filtered ``pink beam'' radiation.\cite{rau2001recent} Condition (iii) can be expressed by the critical propagation distance $z_c=(2a)^2/\lambda$, with pixel size $a$ and X-ray wavelength $\lambda$. For the brain, the \SI{7}{\centi\meter} propagation distance is well within $z_c = $ \SI{137}{\centi\meter}. The kidney had \SI{9}{\centi\meter} distance and $z_c=$ \SI{9.4}{\centi\meter}. Therefore, the SR for both measurements will not be limited by propagation effects.\cite{weitkamp2011ankaphase} Both measurements showed edge enhancement at the outer boundaries of the sample.

The initial SR of the kidney reconstruction was around 6 pixels while the initial SR of the brain measurement was near the Nyquist limit (measured with the method proposed in Ref.\ \onlinecite{mizutani2016method}.) These SR values are relatively consistent with the expected resolution of the detection system according to the formalism developed by Koch \emph{et al.} in Ref.\ \onlinecite{koch98}. If the attenuation lengths in LuAG (\SI{40}{\micro\meter} for \SI{20}{\kilo\electronvolt} and \SI{60}{\micro\meter} for \SI{23}{\kilo\electronvolt}) are considered as effective scintillator depths, the values thus obtained for the expected resolution in terms of the full width of the line-spread function containing 90\% of the intensity response are \SI{9}{\micro\meter} (i.e.\ approximately 2 pixels) for the brain samples and \SI{5}{\micro\meter} (approximately 5 pixels) for the kidney measurements.

Both the Paganin and Gaussian filters can be described as a convolution of the transmission projection $T(x,y) = I(x,y)/I_0(x,y)$. The Paganin filtered projection is given by
\begin{equation} \label{eq_pag}
P(x,y) = \mathcal{F}^{-1} \bigg\{ \frac{1}{1+\frac{\lambda z}{4 \pi}\frac{\delta}{\beta} (u^2+v^2)} \times \mathcal{F}\{ T(x,y) \} \bigg\},
\end{equation}
where $\mathcal{F}$ ($\mathcal{F}^{-1}$) denotes a two-dimensional (inverse) Fourier transform, $u$ and $v$ are the Fourier-space coordinates (in units of inverse pixels) dual to $x$ and $y$, and $z$ the propagation distance.\cite{paganin2002simultaneous,weitkamp2011ankaphase} The Gaussian filtered projections are interpreted as software-blurred transmission projections and are defined by
\begin{equation} \label{eq_gauss}
G(x,y) = \mathcal{F}^{-1} \bigg\{ e^{-2\pi^2\sigma^2(u^2+v^2)} \times \mathcal{F}\{ T(x,y) \} \bigg\},
\end{equation}
where $\sigma$ is the standard deviation of the real-space Gaussian filter (in pixels). Thus, Paganin filtering is multiplication with a Lorentzian in Fourier space, while Gaussian filtering is multiplication with a Gaussian in Fourier space. Taking the logarithm of Eq.\ \ref{eq_pag} and multiplying by $\delta/2\beta$ provides a projected phase shift map, while taking the logarithm of Eq.\ \ref{eq_gauss} and multiplying by $1/(2ka)$ provides the projected $\beta$ map.

For this study, a filtered back-projection with the standard Ram-Lak filter was used for tomographic reconstruction. All analysis steps were implemented in Matlab (The MathWorks, Inc., Natick, USA).

CNR and SR were measured in the reconstructed slices. CNR is defined as $|\mu_1 - \mu_2|/\sqrt{\sigma_1^2 + \sigma_2^2}$ based on the mean ($\mu$) and standard deviation ($\sigma$) within equal-sized, uniform regions of interest (see green and yellow boxes of Figures \ref{fig_slices_brain} and \ref{fig_slices_kidney}). For reference, the difference of the linear attenuation coefficient between these regions of interest is \SI{1.33}{\per\meter} (\SI{43.1}{\per\meter}) for the brain (kidney) sample. The SR was measured with the method proposed by Mizutani \emph{et al.}\cite{mizutani2016method} that does not require a noise criterion. Large values of SR correspond to low spatial resolution. Here, SR is normalized by pixel size.

We define the image quality factor (QF) as
\begin{equation}
\textrm{QF} = \frac{\tanh(\textrm{CNR})}{\textrm{SR}/\textrm{SR}_0},
\end{equation}
where SR$_0=2$ pixels is the Nyquist limit for the minimum SR. Here, QF is used as a image quality metric that is maximized at QF $=1$ for Nyquist-limited spatial resolution and CNR $\rightarrow \infty$. The hyperbolic tangent provides diminishing increases in QF for large CNR. 

Fig.\ \ref{fig_slices_brain} shows characteristic reconstructed slices of the cerebellum sample after Paganin and Gaussian filtering with increasing kernel sizes. The unfiltered reconstruction is given in the top row for reference. Paraffin (p, black), molecular layer (m, darker gray), and granular layer (g, lighter gray) can be identified in all reconstructions. With increasing filter size, both CNR and SR values increase. Each row has approximately equal spatial resolution. The largest QF is achieved with Gaussian filtering (d), though both filters significantly improve image quality (d,e,f). At low (high) values of SR, Gaussian (Paganin) produces higher QF images. It should be noted that all Gaussian filters produced accurate $\beta$ values, while only the Paganin filter with \db$=10^3$ gave accurate $\delta$ as confirmed by previous experiments.\cite{bikis2019sensitivity}

Fig.\ \ref{fig_slices_kidney} shows characteristic reconstructed slices and zooms of the mouse kidney, including the unfiltered datasets (left). Tubular lumina (background, black), tubular tissue (darker gray), and contrast agent in the vascular lumina (lighter gray) can all be identified, particularly after filtering. The highest QF image is produced with Gaussian filter of \sig\ $=2$ pixels. For all SR values, the QF of Gaussian filtering is greater or equal to Paganin filtering. Measured $\beta$ does not depend on the Gaussian filter, while $\delta$ is linear to the \db\ of the Paganin filter. The correct \db\ was unknown for this sample. 

\begin{figure} 
\includegraphics[scale=0.6]{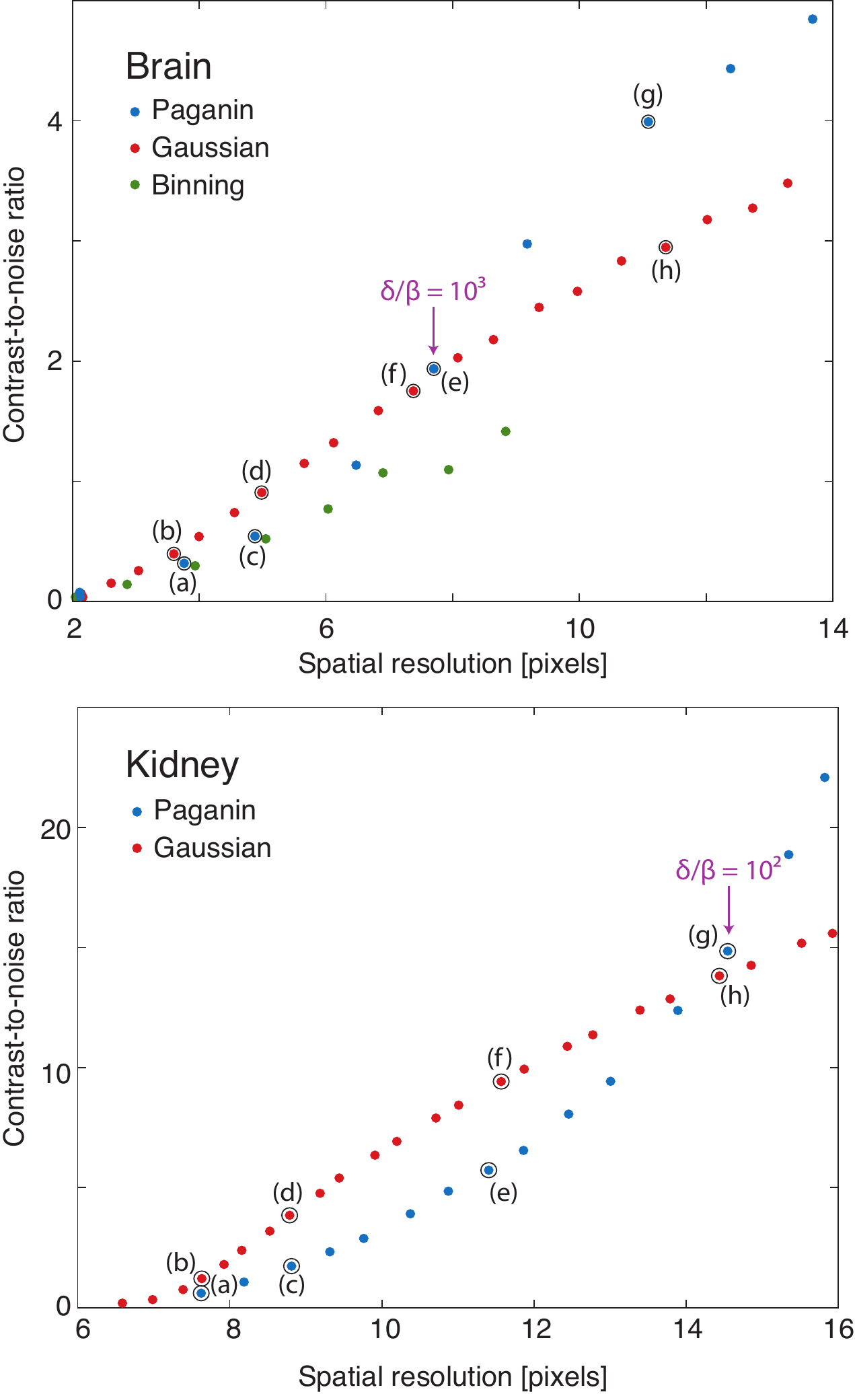}%
\caption[caption with footnote]{\label{fig_sr_vs_cnr} Measured CNR and SR (in pixels) for the reconstructed slices of human brain (top) and mouse kidney (bottom) after Paganin phase retrieval (orange) and Gauss filtering (blue). Binning up to $8\times8$ pixels is also given for the brain for reference (green). Both filters show a sigmoidal relationship between CNR and SR until at very large SR uniform regions of interest (ROIs) become difficult to select. Slices shown in Figures \ref{fig_slices_brain}(a-h) and \ref{fig_slices_kidney}(a-j) are indicated.}%
\end{figure}

The CNR is plotted against the SR for both samples in Fig.\ \ref{fig_sr_vs_cnr}. At small values of SR, Gaussian filtering provides greater CNR. At larger values of SR, Paganin filtering produces greater CNR. For the brain, the crossover is at \db\ $=10^3$, where accurate $\delta$ values are given by Paganin filtering. For the kidney, the crossover point was at \db\ $=10^2$. The QF in these larger filter kernel regions is lower because CNR $>1$ and therefore the increasing SR is the dominant factor. For the brain specimen, the results of data binning\cite{thurner2004optimization,thalmann2017single} are given, though CNR is lower than for Gaussian filtering for any SR. For convenience, the software binned data were nearest-neighbor interpolated back to their original size prior to reconstruction to have equal pixel size to the other filtered datasets.

The maximum QF reached for the brain (kidney) dataset was $0.29$ ($0.24$) for Gaussian filtering, $0.25$ ($0.21$) for Paganin filtering, and $0.23$ for software binning. The maximum QF for the kidney dataset was achieved at $\delta/\beta=5$ while for the brain it was for $\delta/\beta=500$ and $1000$. For context, the full width at half maximum of the Paganin filter for the kidney dataset at $\delta/\beta=5$ is nearly equal to that in the brain dataset at $\delta/\beta=100$ (from Eq.\ \ref{eq_pag} and the measurement parameters in Table \ref{tab_acq_params}.) Still, the $\delta/\beta$ optimizing QF for the Paganin filter has no inherent physical meaning. The trade-off between SR and CNR for Paganin filtering will depend on the noise and spatial resolution of the imaging system rather than the accuracy of the phase retrieval.

It should be noted that while software binning shows smaller QF compared to Gaussian filtering for any SR, it provides the benefit of reducing the three-dimensional dataset's size by the cube of the binning factor. 
For this study, a two-dimensional Gaussian filter was applied to the projections to make an analogy with the Paganin filter. Three-dimensional Gaussian filtering of the reconstructed data provided similar results with maximum quality factor within 1\% of two-dimensional filtering, although previous studies have shown that binning prior to reconstruction provides benefits over binning reconstructed data.\cite{thurner2004optimization}
The combination of Gaussian filtering before binning can provide the benefits of both strategies. Mean and median filters also provided lower QF at any SR compared with Gaussian filtering, therefore they are not presented here. Further studies should determine how Gaussian filtering compares with more advanced filters such as bilateral filtering.\cite{tomasi1998bilateral,manduca2009projection}

For reference, the difference in linear attenuation coefficient between the regions of interest is \SI{1.33}{\per\meter} in Fig.\ \ref{fig_slices_brain} and \SI{43.1}{\per\meter} in Fig.\ \ref{fig_slices_kidney}. The density resolution varies with filter kernel size and can be calculated considering the CNR. Interpreting density resolution for the Paganin phase retrieval is more challenging. For a comparison with other techniques such as grating interferometry, several other factors must be considered in addition to the trade-off between CNR and SR.\cite{bikis2019sensitivity}

Future studies are needed to determine the effects of noise level of the projections, sample composition, and propagation distance. Previous results have shown that certain single-distance phase retrieval results are robust against noise,\cite{boone2012improved} though the initial noise level will limit the maximum achievable CNR and the slope of the SR vs. CNR curve. We expect increasing the propagation distance will benefit the performance of the Paganin filter over Gaussian filtered absorption, although the latter may still perform better than previously expected. Characterizing the beam coherence\cite{lin2003measurement,pfeiffer2005shearing,kashyap2015two} will help quantify the contributions of phase information versus low pass filtering to CNR gains. Paganin's approach likely proves more advantageous when absorption is negligible.

Gaussian filtering produces higher quality images at high spatial resolution (low SR values) for the two biomedically relevant samples considered here. Being model-independent, it also allows for the quantitative interpretation of gray values. Paganin filtering, though more commonly used, only outperformed Gaussian filtering at low spatial resolution where CNR was already large. These results suggest that for certain medically relevant specimens, the density resolution improvements from phase retrieval can be matched or exceeded by suitably low-pass filtered absorption measurements. These results are especially meaningful for samples with non-neglible absorption. We conclude that both the SR and CNR must be considered when comparing the quality of tomography data.

\begin{acknowledgments}
The authors thank S.\ Marathe, K. Wanelik, and A.\ Bodey, Diamond Light Source I13-2, for support during the beam time. The authors thank C.\ Bikis, University of Basel, and S.\ Theocharis, National and Kapodistrian University of Athens, for providing the brain specimen. The project was supported by Diamond Light Source projects MT 19829-1 and MT 20746. T.W.\ acknowledges support from the French National Research Agency through the EQUIPEX investment program, project NanoimagesX, Grant No.\ ANR-11-EQPX-0031. The authors acknowledge support from the Swiss National Science Foundation through NCCR Kidney.CH as well as Project Nos.~185058 and 153523.
\end{acknowledgments}


%

\end{document}